\documentclass[aps,prd,twocolumn,amsmath,amssymb]{revtex4-1}
\usepackage{mathtools}
\usepackage{graphicx}
\usepackage{subfigure}
\usepackage{epstopdf}
\usepackage{color}
\usepackage{multirow}
\usepackage{setspace}
\usepackage{overpic}
\usepackage[bookmarksnumbered, pdfstartview=FitH,colorlinks,urlcolor=blue, citecolor=blue,linkcolor=blue] {hyperref}
\usepackage{lineno}
\usepackage{bm}
\usepackage{rotating}
\usepackage[utf8]{inputenc}
\usepackage{morefloats}
\usepackage{hyperref}
\usepackage{braket}

\let\oldequation\equation
\let\oldendequation\endequation

\renewenvironment{equation}
  {\linenomathNonumbers\oldequation}
  {\oldendequation\endlinenomath}

\newcommand{\BESIIIorcid}[1]{\href{https://orcid.org/#1}{\hspace*{0.1em}\raisebox{-0.45ex}{\includegraphics[width=1em]{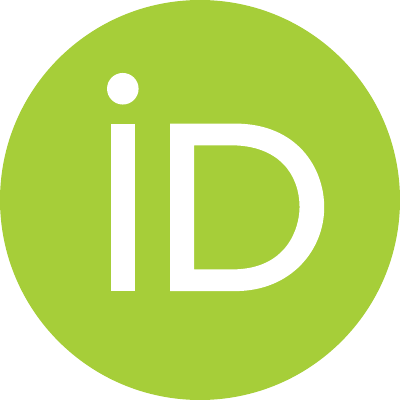}}}}

\begin{document}
%\linenumbers

\title{\boldmath Measurements of the inclusive Branching fractions \\of the $D_s^+ \to \eta (\eta^\prime) X$ decays}
\author{M.~Ablikim \emph{et al.}}
\thanks{Full author list given at the end of the article.}
\collaboration{BESIII Collaboration}
\noaffiliation
\date{\today}

\begin{abstract}
Using $e^+ e^-$ collision data collected with the BESIII detector at center-of-mass energies between $4.128$ and $4.226$ GeV, corresponding to an integrated luminosity of $7.33~{\rm fb}^{-1}$,
the inclusive branching fractions of the decays $D_s^+ \to \eta X$ and $D_s^+ \to \eta^\prime X$ ($X$ denotes any possible particles) are measured to be $(30.72 \pm  0.90 \pm 0.46)\%$ and $(12.30 \pm  0.94 \pm 0.37)\%$, where the first uncertainties are statistical and the second systematic.  
The results are consistent with the previous CLEO measurements, with precision improved by factors of 2.8 and 1.9 for $D_s^+ \to \eta X$ and $D_s^+ \to \eta^\prime X$, respectively.
\end{abstract}

\maketitle
\section{Introduction}
The $D_s^+$ meson is the ground state of charmed-strange mesons.
Precision measurements of its branching fractions (BFs) provide a more complete understanding of $D_s^+$ decay dynamics
and serve as essential inputs for normalizing $B$ and $B_s$ decays, constraining systematic uncertainties, and improving background modeling.
However, the sum of the measured exclusive BFs of $D_s^+$ decays is only $(85.7\pm3.7)\%$~\cite{PDG}, leaving significant room for unobserved decay modes.
The $D_s^+$ decays are dominated by Cabibbo-favored transitions to final states containing $\eta$ and $\eta^\prime$, which currently account for over one third of all $D_s^+$ decays.
Based on the exclusive decay modes listed in Tables~\ref{Tab:BF1} and~\ref{Tab:BF2}~\cite{PDG}, the sums of the BFs for all known $D_s^+$ decays that include an $\eta$ or $\eta^{\prime}$ are $\sum_i \mathcal{B}(D_s^+ \to \eta X_i) = (25.40 \pm 0.20)\%$ and $\sum_i \mathcal{B}(D_s^+ \to \eta^{\prime} X_i) = (11.97 \pm 0.21)\%$, respectively.
While these sums agree with the current Particle Data Group~(PDG) inclusive BFs of $(29.9 \pm 2.8)\%$ and $(10.3 \pm 1.4)\%$~\cite{PDG}, their significantly smaller uncertainties motivate high-precision inclusive measurements.

The inclusive BFs of $D^+_s\to \eta X$ and $D^+_s\to \eta^\prime X$  were  previously measured by the CLEO and BESIII collaborations. Using 586 pb$^{-1}$ of $e^+e^-$ collision data taken at the center-of-mass energy $\sqrt{s}=4.170$ GeV, CLEO reported $\mathcal{B}(D_s^+ \to \eta X) = (29.90 \pm 2.20 \pm 1.70)\%$ and $\mathcal{B}(D_s^+ \to \eta^\prime X) = (11.70 \pm 1.70 \pm 0.70)\%$~\cite{Ref:CLEO2009}, respectively. BESIII measured $\mathcal{B}(D_s^+ \to \eta^\prime X) = (8.8 \pm 1.8 \pm 0.5)\%$ with 482 pb$^{-1}$ of $e^+e^-$ collision data taken at $\sqrt{s}=4.009$ GeV~\cite{BESIII:2015rrp}. All previous measurements are dominated by substantial statistical uncertainties. 
In this paper, we measure the inclusive BFs of $D^+_s\to \eta X$ and $D^+_s\to \eta^\prime X$ using $e^+e^-$ collision data, corresponding to an integrated luminosity of 7.33 fb$^{-1}$, collected with the BESIII detector at $\sqrt{s} = 4.128$-$4.226~\rm{GeV}$. 
Charge-conjugate states are implied throughout this paper. 
The $e^+e^- \to D_s^{*\pm} D_s^{\mp} (\to \gamma D_s^+ D_s^-)$ process allows the application of the double-tag (DT) method~\cite{PhysRevLett.56.2140}
to acquire signal samples. 
In this method, since the $D_s^+$ and $D_s^-$ mesons are produced in pairs, the hadronic $D_s^-$ mesons are fully reconstructed on the tag side,
while the signal candidates are identified by analyzing the remaining charged tracks on the recoil side that are not used in the tag-side reconstruction.
Events where a tag meson is found are referred to as single-tag (ST) events, and events where a signal decay is identified on the recoil side in
addition to the tag meson are referred to as double-tag (DT) events.
The ST candidates are fully reconstructed via any of the four hadronic decay channels:  \(D^-_s \to K_S^0 K^-\), \(K^+ K^- \pi^-\), \(K^+ K^- \pi^- \pi^0\), or \(K_S^0 K^+ \pi^- \pi^-\).
The \(e^+e^- \to D_s^{*\pm} D_s^{\mp}\) process is used because of its high cross section, which is about twenty times larger than that
of the \(e^+e^- \to D_s^+ D_s^-\)  process in this energy range~\cite{PhysRevD.80.072001}.
%------------------------------------------------
\begin{table}[htbp]
	\caption{The BFs of known exclusive $D_s^+ \to \eta X$ decays from the PDG~\cite{PDG}. The listed uncertainties are the combined statistical and systematic  uncertainties. The BF of $D_s^+ \to \eta X$ via $\eta^\prime$ is calculated by all the sum of BFs of $D_s^+ \to \eta^\prime X$ listed in Table~\ref{Tab:BF2} multiplied by the BFs of $\eta^\prime \to \eta \pi^+ \pi^-$ and $\eta^\prime \to \eta \pi^0 \pi^0$~\cite{PDG}. To avoid double counting, the BF of $D_s^+ \to \eta^\prime \pi$ ( $\eta^\prime \to \pi^+ \pi^- \eta$) is excluded.}
	\begin{tabular}{lr@{$\pm$}l}
		\hline
		Decay mode                       		&\multicolumn{2}{c}{BF~(\%)}       	\\
		\hline
		$D_s^+ \to \eta e^+ \nu_e$         		                &$2.27$ & $0.06$                   	\\       
		$D_s^+ \to \eta \mu^+ \nu_{\mu}$   		                &$2.24$ & $0.07$                   	\\       
		$D_s^+ \to \eta \pi^+$             		                &$1.686$ & $0.027$                 	\\       
		$D_s^+ \to \eta \pi^+ \pi^0$       		                &$9.10$ & $0.17$                  	\\             
		$D_s^+ \to \eta K^+$               		                &$0.176$ & $0.008$                 	\\       
		$D_s^+ \to \eta \pi^+ \pi^+ \pi^-$   	                &$3.08$ & $0.08$    		         \\       
		$D_s^+ \to \eta X$ via $\eta'$       	                &$6.08$ & $0.14$    		         	\\       
		$D_s^+ \to \eta \omega \pi^+$   		                &$0.54$ &$0.13$		              \\     
		$D_s^+ \to \pi^+ \phi,~\phi \to \gamma \eta$            &$0.06 $ & $ 0.01$              \\        
		$D_s^+ \to \rho^+ \phi,~\phi \to \gamma \eta$           &$0.07 $ & $ 0.01$               	\\        
		$D_s^+ \to e^+ \nu_e \phi,~\phi \to \gamma \eta$                              &     $0.03 $ & $ 0.01$            	\\        
		$D_s^+ \to \mu^+ \nu_\mu \phi,~\phi \to \gamma \eta$                          &     $0.03$ & $ 0.01$            	\\        
		\hline
		Sum                              		                &$25.40$ & $0.20$                  	\\
		\hline
	\end{tabular}
	\label{Tab:BF1}
\end{table}

%------------------------------------------------
\begin{table}[htbp]
	\caption{The BFs of known exclusive $D_s^+ \to \eta^\prime X$ decays from the PDG~\cite{PDG}. The listed uncertainties are the combined statistical and systematic uncertainties.}
	\begin{tabular}{lr@{$\pm$}l}
		\hline
		Decay mode                         			&\multicolumn{2}{c}{BF~(\%)}        	\\
		\hline
		$D_s^+ \to \eta^\prime e^+ \nu_e$        	&$0.81$ & $0.04$    \\       
		$D_s^+ \to \eta^\prime \mu^+ \nu_{\mu}$  	&$0.80$ & $0.06$    \\       
		$D_s^+ \to \eta^\prime \pi^+$            	&$3.95$ & $0.08$    \\       
        $D_s^+ \to \eta^\prime \pi^+ \pi^0$         &$6.14$ & $0.18$    \\       
		$D_s^+ \to \eta^\prime K^+$              	&$0.268$ & $0.024$  \\     

        \hline
		Sum                              			&$11.97$ & $0.21$   \\
   		\hline
	\end{tabular}
	\label{Tab:BF2}
\end{table}

%------------------------------------------------------------------------------
\section{Detector and data sets}

The BESIII detector~\cite{Ablikim:2009aa} records symmetric $e^+e^-$ collisions 
provided by the BEPCII storage ring~\cite{Yu:IPAC2016-TUYA01}
in the center-of-mass energy range from 1.84 to 4.95~GeV,
with a peak luminosity of $1.1 \times 10^{33}\;\text{cm}^{-2}\text{s}^{-1}$ 
achieved at $\sqrt{s} = 3.773\;\text{GeV}$. 
BESIII has collected large data samples  at $\sqrt{s} = 1.84$-$4.95$ GeV. The cylindrical core of the BESIII detector covers 93\% of the full solid angle and consists of a helium-based
 multilayer drift chamber~(MDC), a time-of-flight
system~(TOF), and a CsI(Tl) electromagnetic calorimeter~(EMC),
which are all enclosed in a superconducting solenoidal magnet
providing a 1.0~T magnetic field.
The solenoid is supported by an
octagonal flux-return yoke with resistive plate counter muon
identification modules interleaved with steel. 
%The acceptance of charged particles and photons is 93\% over $4\pi$ solid angle. 
The charged-particle momentum resolution at $1~{\rm GeV}/c$ is
$0.5\%$, and the 
${\rm d}E/{\rm d}x$
resolution is $6\%$ for electrons
from Bhabha scattering. The EMC measures photon energies with a
resolution of $2.5\%$ ($5\%$) at $1$~GeV in the barrel (end cap)
region. The time resolution in the plastic scintillator TOF barrel region is 68~ps, while
that in the end cap region was 110~ps.  The end cap TOF
system was upgraded in 2015 using multigap resistive plate chamber
technology, providing a time resolution of
60~ps,
which benefits 84\% of the data used in this work~\cite{etof,etof1,etof2}.
 
The data samples used in this paper correspond to a total integrated luminosity of 7.33 fb$^{-1}$, with the luminosity breakdown by individual center-of-mass energies given in Table~\ref{energe}.
Considering the low statistics of some data samples, we group them together as follows: $4.128$-$4.157$~GeV, $4.178$~GeV, $4.189$-$4.219$~GeV and $4.226$~GeV.

\begin{table}[!htbp]
\renewcommand\arraystretch{1.25}
  \caption{The integrated luminosities~($\mathcal{L}_{\rm int}$) and the requirements on $M_{\rm rec}$ for various center-of-mass energies~\cite{s1,l}. 
   The first and second uncertainties are statistical and systematic, respectively. 
   The definition of $M_{\rm rec}$ is given in Eq.~(\ref{eq:mrec}). 
   The integrated luminosities for the data samples of $\sqrt{s} = 4.128~\rm{GeV}$ and $\sqrt{s} = 4.157~\rm{GeV}$ are given by online monitoring information.}
 \begin{tabular}{c c c}
 \hline
 $\sqrt{s}$~(GeV) & $\mathcal{L}_{\rm int}$~(pb$^{-1}$) & $M_{\rm rec}$~(GeV/$c^2$)\\
 \hline
  4.128 &  401.5                      & [2.060, 2.150] \\
  4.157 &  408.7                      & [2.054, 2.170] \\
  4.178 & $3189.0\pm0.2\pm31.9$ & [2.050, 2.180] \\
  4.189 &  $570.0\pm0.1\pm2.2$  & [2.048, 2.190] \\
  4.199 &  $526.0\pm0.1\pm2.1$  & [2.046, 2.200] \\
  4.209 &  $572.1\pm0.1\pm1.8$  & [2.044, 2.210] \\
  4.219 &  $569.2\pm0.1\pm1.8$  & [2.042, 2.220] \\
  4.226 & $1100.9\pm0.1\pm7.0$  & [2.040, 2.220] \\
  \hline
 \end{tabular}

    \label{energe}
\end{table}

Monte Carlo (MC) simulated data samples produced with a {\sc
geant4}-based~\cite{geant4} software package, which
includes the geometric description of the BESIII detector and the
detector response, are used to determine detection efficiencies
and to estimate backgrounds. The simulation models the beam
energy spread and initial-state radiation (ISR) in the $e^+e^-$
annihilations with the generator {\sc
kkmc}~\cite{ref:kkmc}.
The inclusive MC sample includes the production of open charm
processes, the ISR production of vector charmonium(-like) states,
and the continuum processes incorporated in {\sc
kkmc}~\cite{ref:kkmc}.
This MC sample is used to optimize selection criteria, investigate distributions of signal and background processes, and determine the efficiencies of our selection criteria. 
All particle decays are modeled with {\sc
evtgen}~\cite{ref:evtgen} using BFs 
either taken from the
PDG~\cite{PDG}, when available,
or otherwise estimated with {\sc lundcharm}~\cite{ref:lundcharm}.
Final-state radiation from charged final state particles is incorporated using {\sc
photos}~\cite{photos2}.

%------------------------------------------------------------------------------

\section{Event selection}
\label{Event-selection}

The $D_s^{-}$ candidates are constructed from combinations of $\pi^{\pm}$, $\pi^0$, $K^{\pm}$, $K_S^0$, $\eta$, $\eta^\prime$, and $\gamma$ candidates.

Charged tracks detected in the MDC are required to be within a polar angle ($\theta$) range of $\vert\!\cos\theta\vert<0.93$, where $\theta$ is defined with respect to the $z$ axis,
which is the symmetry axis of the MDC. 
For charged tracks not originating from $K_S^0$ decays, the distance of closest approach to the interaction point 
must be less than 10\,cm
along the $z$ axis, $|V_{z}|$,  
and less than 1\,cm
in the transverse plane, $|V_{xy}|$. 
Particle identification~(PID) for charged tracks combines measurements of d$E$/d$x$ and the flight time in the TOF to form likelihoods $\mathcal{L}(h)~(h=K,\pi)$ for each hadron $h$ hypothesis.
The charged kaons and pions are identified by comparing the likelihoods for the kaon and pion hypotheses, $\mathcal{L}(K)>\mathcal{L}(\pi)$ and $\mathcal{L}(\pi)>\mathcal{L}(K)$, respectively.

Each $K_{S}^0$ candidate is reconstructed from two oppositely charged tracks satisfying $|V_{z}|<$ 20~cm.
The two charged tracks are assigned
as $\pi^+\pi^-$ without imposing further PID criteria. They are constrained to
originate from a common vertex and are required to have an invariant mass
within $|M_{\pi^{+}\pi^{-}} - m_{K_{S}^{0}}|<$ 12~MeV$/c^{2}$, where
$m_{K_{S}^{0}}$ is the known $K^0_{S}$ mass~\cite{PDG}.

Photon candidates are identified using isolated showers in the EMC.  The deposited energies of each shower must be more than 25~MeV in the barrel region ($|\cos \theta|< 0.80$) and more than 50~MeV in the end cap region ($0.86 <|\cos \theta|< 0.92$).  
To exclude showers that originate from
charged tracks,
the angle subtended by the EMC shower and the position of the closest charged track at the EMC
must be greater than  10 degrees as measured from the interaction point. 
To suppress electronic noise and showers unrelated to the event, the difference between the EMC time and the event start time is required to be within 
[0, 700]\,ns.

The $\pi^0$ and $\eta$ candidates are reconstructed from photon pairs with invariant masses in the ranges $[0.115,~0.150]$~GeV/$c^{2}$ and $[0.360,~0.740]$~GeV/$c^{2}$, respectively.
In order to improve the invariant mass resolutions, at least one photon is required to come from the barrel region of the EMC.
Furthermore, the $\pi^0$ candidates are constrained to the known $\pi^0$ mass~\cite{PDG} via a mass constraint  kinematic fit to improve their energy and momentum resolution and the \(\chi^{2}\) of this kinematic fit is  required to be less than 30. 

The $\eta^{\prime}$ candidates are reconstructed from $\pi^{+}\pi^{-}\eta$ combinations, with the $\eta$ decaying to two photons. 
The invariant mass of two photons is required to be in the range of [0.50, 0.57] GeV/$c^2$. 
Furthermore, 
a kinematic fit constraining the invariant mass of the selected photon pair to the known $\eta$ mass, $m_{\eta}$~\cite{PDG}, is performed, 
and the $\chi^2$ of the kinematic fit is required to be less than 30.

The invariant masses of $D_s^{-}$ candidates~($M_{\rm tag}$) are required to be in the range of $[1.88,~2.06]~{\rm GeV}/c^2$. 
The recoil mass $M_{\rm rec}$ is defined as
\begin{eqnarray}
\small
\begin{aligned}
	\begin{array}{l r}
	M_{\rm rec}^2 c^4  = \left (\sqrt{s} - \sqrt{|\vec{p}_{D_{s}^{-}}|^{2} c^2+m^2_{D_{s}^{-}}c^4} \right)^{2} - \left|\vec{p}_{D_{s}^{-}} \right| ^{2} c^2 \; ,
		\end{array}
\end{aligned}
 \label{eq:mrec}
\end{eqnarray}
%----------------------------------
where $\vec{p}_{D_{s}^{-}}$ is the three-momentum of the $D_{s}^{-}$ candidate in the $e^+e^-$ center-of-mass frame 
and $m_{D_{s}^{-}}$ is the known $D_{s}^{-}$ mass~\cite{PDG}. 
The requirements on $M_{\rm rec}$ for all energy points are given in Table~\ref{energe}. 
The photon from $D_s^{*\pm}$ decays is selected by requiring the mass of the other $D_s^{\mp}$, recoiling against the $D_s^{\pm}$ and the photon, to be consistent with the known $D_s^+$ mass~\cite{PDG}.
If there are multiple ST candidates,  the candidate with the $M_{\rm rec}$ closest to the known $D_{s}^{*\pm}$  mass~\cite{PDG} is kept.

To select signal events with $D_s^+ \to \eta(\eta^\prime) X$, we require that the DT events contain an $\eta$ ($\eta^\prime$) candidate among the particles recoiling against the ST candidates.
All DT candidates are retained. 
To improve the resolution of the mass calculation, the $\eta$ signal yield is then determined by fitting the $M_{\gamma\gamma}$ distribution (denoted as $M_{\eta}$), and the $\eta^\prime$ signal yield is determined by fitting the $M_{\pi^+ \pi^- \gamma\gamma} - M_{\gamma\gamma} + m_{\eta}$ distribution (denoted as $M_{\eta^\prime}$), where $M_{\gamma\gamma}$ is the invariant mass of the $\gamma\gamma$ combinations, $M_{\pi^+ \pi^- \gamma\gamma}$ is the invariant mass of the $\pi^+\pi^-\gamma\gamma$ combinations. %, and $m_{\eta}$ represents the known $\eta$ mass~\cite{PDG}.

%-----------------------------------------------------------------------------

\section{ST and DT yields}
\label{section:yields}
\begin{figure*}[htpb]
	\centering
  		\includegraphics[width=7cm]{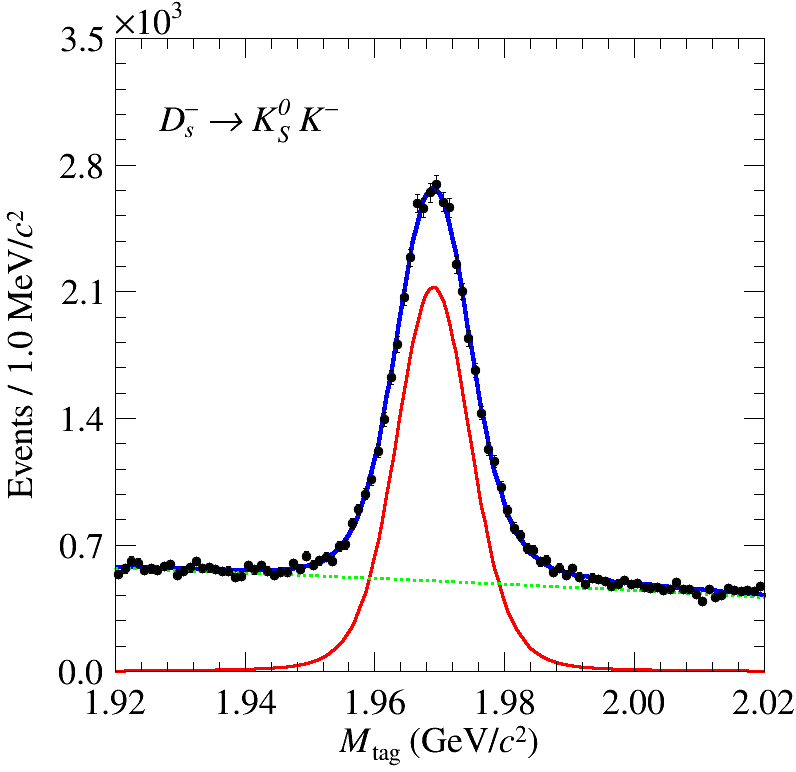}
		\includegraphics[width=7cm]{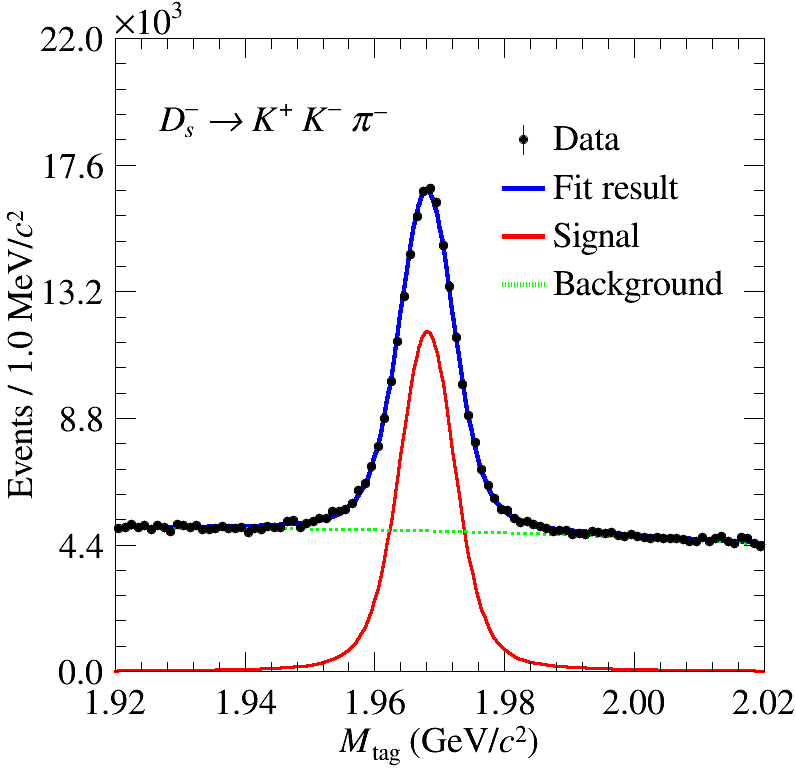}\\
  		\includegraphics[width=7cm]{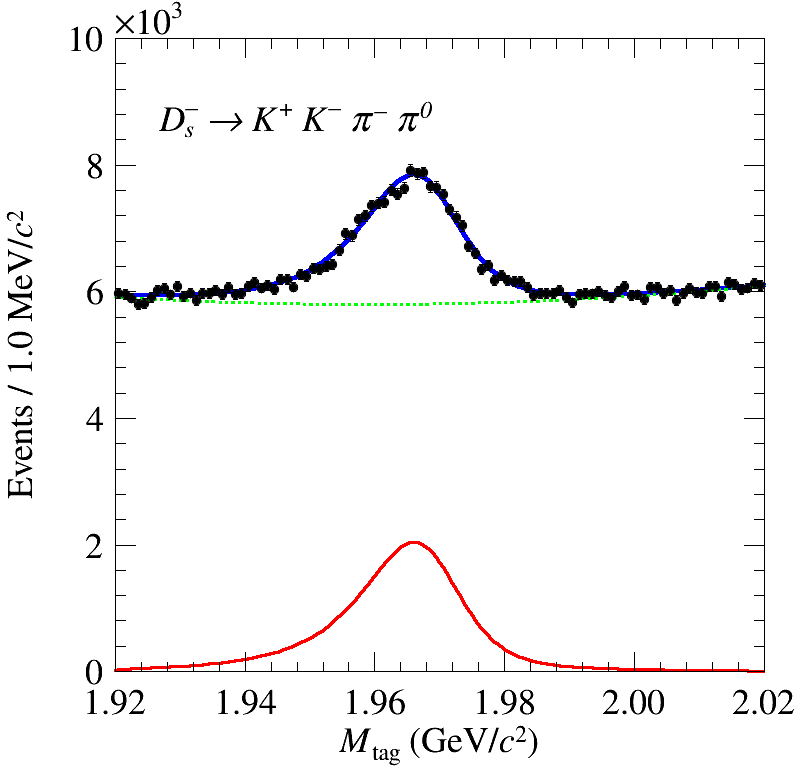}
		\includegraphics[width=7cm]{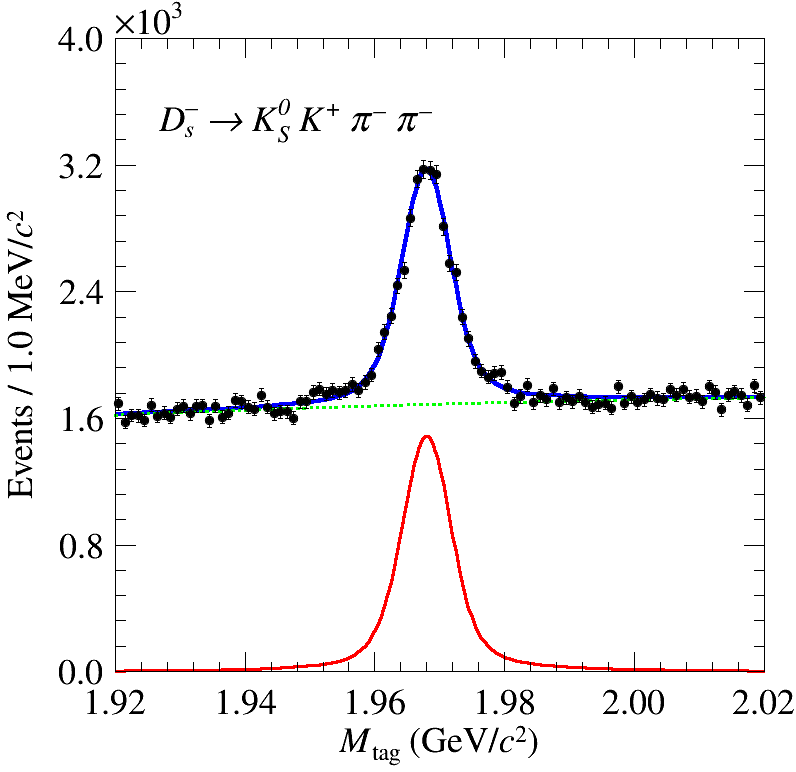}
	\caption{Fits to the distributions of $M_{\rm tag}$ for each tag mode from the data sample are performed to obtain the ST yields at $\sqrt{s}=4.178$~GeV. The dots with error bars are data. The total fit results are shown as the solid blue curves. The background and signal components are depicted as the dashed green and solid red curves, respectively.}
	\label{Fig:ST} 
\end{figure*}

The ST yields of each tag mode are determined from the fits to individual $M_{\rm tag}$ distributions.
In the fits, the signal shape is modeled by the MC-simulated shape convolved with a Gaussian function to account for the detector resolution, and the background shape is parameterized as a second-order Chebyshev polynomial, as shown in Fig.~\ref{Fig:ST}.
In the fit of $D^-_s \to K_S^0 K^-$, the MC-simulated shapes of $D^- \to K_S^0 \pi^-$ are added to the background polynomial function accounting for the peaking background and the ratios of both the peaking background and the polynomial function are left as free parameters.
The ST yields in data and the ST efficiencies estimated with the inclusive MC samples are  listed in Tables~\ref{Tab:ST} and \ref{Tab:STeff}, respectively.

\begin{table*}[htbp]
	\caption{ The ST yields $(\times 10^3)$ in each data sample. The uncertainties are statistical.} 

	\begin{tabular}{l r@{$\pm$}l r@{$\pm$}l r@{$\pm$}l r@{$\pm$}l}
		\hline
    Tag mode  &  \multicolumn{2}{c}{$4.128$-$4.157$ GeV} & \multicolumn{2}{c}{$4.178$ GeV} & \multicolumn{2}{c}{$4.189$-$4.219$ GeV} & \multicolumn{2}{c}{$4.226$ GeV}\\
		\hline
		$D^-_s\to K_S^0K^-$              & 6.73 & 0.14   & 31.95 & 0.31   & 19.96 & 0.27   & 6.84 & 0.16  \\
		$D^-_s\to K^+K^-\pi^-$           & 27.67 & 0.28   & 137.14 & 0.61   & 86.92 & 0.53   & 29.54 & 0.34  \\
		$D^-_s\to K^+K^-\pi^-\pi^0$     & 7.46 & 0.40   & 39.34 & 0.80   & 24.69 & 0.69   & 8.08 & 0.48  \\
		$D^-_s\to K_S^0K^+\pi^-\pi^-$   & 2.98 & 0.13   & 15.69 & 0.29   & 9.78 & 0.25   & 3.38 & 0.17  \\
		\hline
	\end{tabular}
	\label{Tab:ST}
\end{table*}

\begin{table*}[htbp]
	\caption{The ST efficiencies (\%). The uncertainties are statistical.} 

	\begin{tabular}{l r@{$\pm$}l r@{$\pm$}l r@{$\pm$}l r@{$\pm$}l}
		\hline
    Tag mode  &  \multicolumn{2}{c}{$4.128$-$4.157$ GeV} & \multicolumn{2}{c}{$4.178$ GeV} & \multicolumn{2}{c}{$4.189$-$4.219$ GeV} & \multicolumn{2}{c}{$4.226$ GeV}\\
		\hline
		$D^-_s\to K_S^0K^-$              & 47.64 & 0.16   & 47.39 & 0.07   & 47.23 & 0.09   & 47.95 & 0.16  \\
		$D^-_s\to K^+K^-\pi^-$           & 40.37 & 0.07   & 39.47 & 0.03   & 39.33 & 0.04   & 39.78 & 0.07  \\
		$D^-_s\to K^+K^-\pi^-\pi^0$      & 10.59 & 0.08   & 10.68 & 0.03   & 10.74 & 0.05   & 10.89 & 0.09  \\
		$D^-_s\to K_S^0K^+\pi^-\pi^-$    & 21.30 & 0.14   & 21.85 & 0.06   & 21.66 & 0.08   & 22.27 & 0.16  \\
		\hline
	\end{tabular}
	\label{Tab:STeff}
\end{table*}

\begin{figure*}[htbp]
  \centering
  \includegraphics[width=7cm]{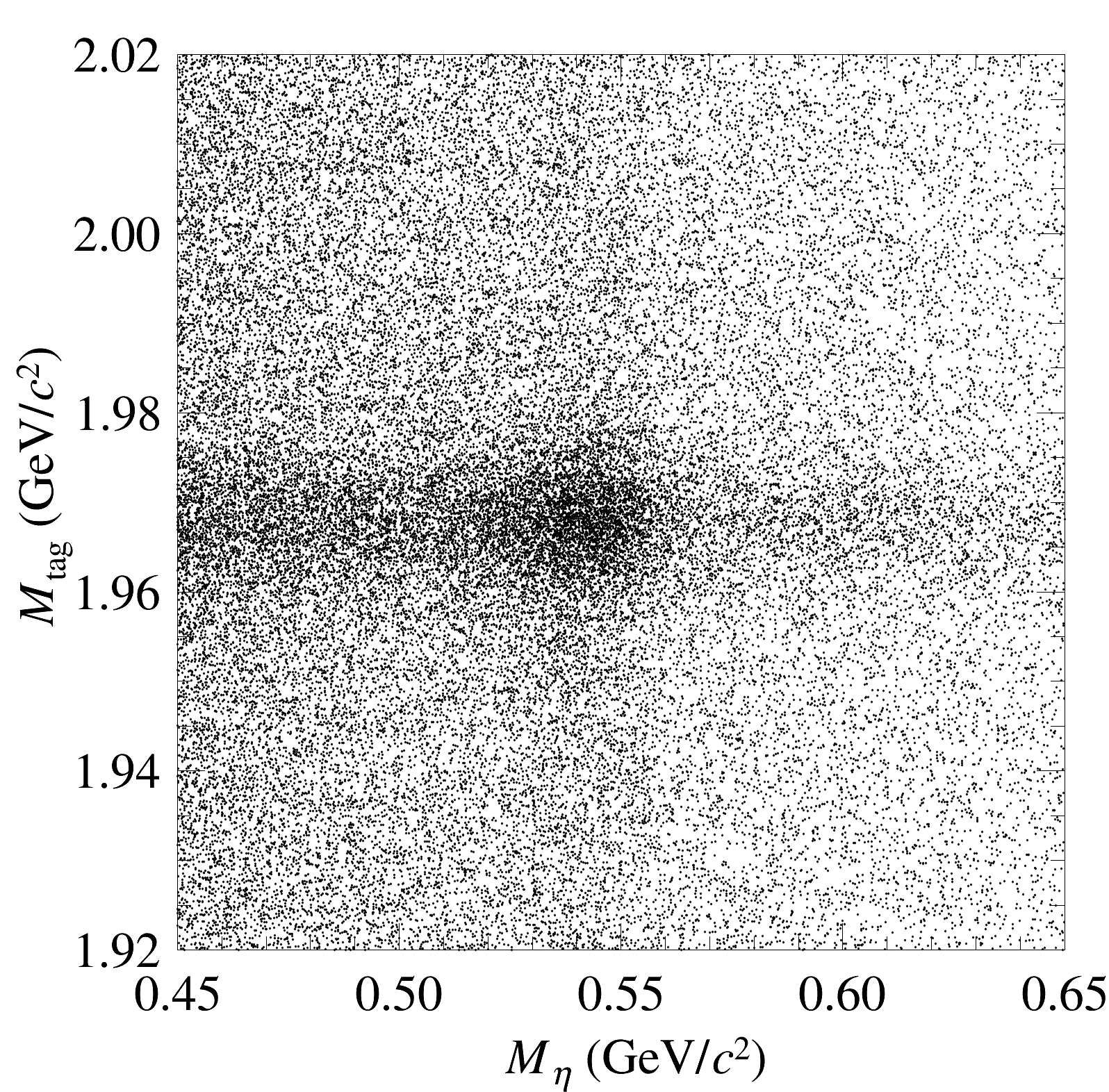}
  \includegraphics[width=7cm]{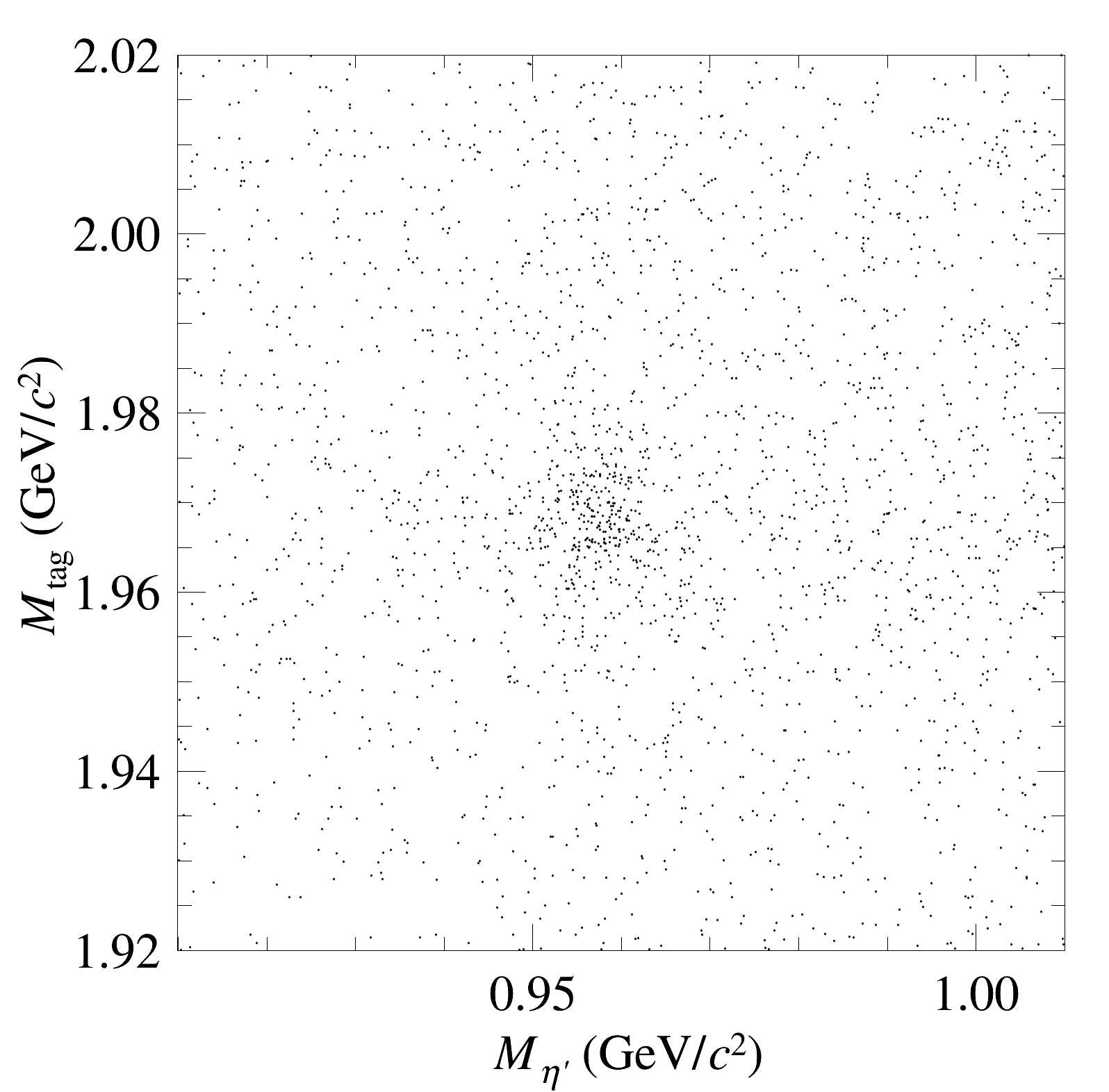}
  \caption{Two-dimensional distributions of $M_{\rm tag}$ versus
$M_{\eta}$ (left) and $M_{\rm tag}$ versus $M_{\eta^\prime}$ (right)
for DT candidates in $D_s^+ \to \eta X$ and $D_s^+ \to \eta^\prime X$ decays in data.}

  \label{Fig:2Ddata}
\end{figure*}

For $D_s^+ \to \eta X$ and $D_s^+ \to \eta^\prime X$, two distinct
types of backgrounds can be clearly distinguished, as illustrated in
Fig.~\ref{Fig:2Ddata}. Backgrounds from non-$D_s^{*\pm}D_s^{\mp}$
events exhibit a peak in the $M_{\eta^{(\prime)}}$ distribution but
show no structure in the $M_{\rm tag}$ distribution. Conversely,
backgrounds originating from $D_s^{\pm}$ decays peak in the
$M_{\rm tag}$ distribution but appear flat in the $M_{\eta^{(\prime)}}$
distribution. Therefore, we perform a two-dimensional (2D) fit on
$M_{\eta}$ ($M_{\eta^\prime}$) versus $M_{\rm tag}$ to determine the
DT yields.

The expected shapes for the signal and background components in the 2D fit include: 

\begin{itemize}
    \item \textbf{Shape1}: the signal shape of the $\eta(\eta^\prime)$ signal side is described by the MC-simulated shape convolved with a Gaussian function,
    \item \textbf{Shape2}: the signal shape of the tag side is described by the MC-simulated shape convolved with a Gaussian function,
    \item \textbf{Shape3}: the background shape of the $\eta(\eta^\prime)$ signal side is described by a third-order Chebyshev polynomial,
    \item \textbf{Shape4}: the background shape of the tag side is described by a second-order Chebyshev polynomial.
\end{itemize}

In the 2D fits, the signal and background shapes are constructed as below:
%------------------------------------------------
\begin{itemize}

	\item \textbf{Signal shape} is constructed as a product of \textbf{Shape1} and \textbf{Shape2},

     \item \textbf{Background I} corresponds to the background that has a peaking structure in the $\eta$ ($\eta^\prime$) mass distributions  but is flat in the tagged $D_{\rm tag}$ distribution. It is defined as the product of \textbf{Shape1} and \textbf{Shape4},

  \item \textbf{Background II} corresponds to the background that has a peaking structure in the tagged $D_{\rm tag}$ distribution but is flat in the $M_{\eta^{(\prime)}}$ distributions. It is defined as the product of \textbf{Shape2} and \textbf{Shape3},
		
	\item \textbf{Background III} represents the non-peaking shape of background, constructed as a product of  \textbf{Shape3} and \textbf{Shape4}.

\end{itemize}

\begin{figure*}[htbp]
	\centering
		\includegraphics[width=7cm]{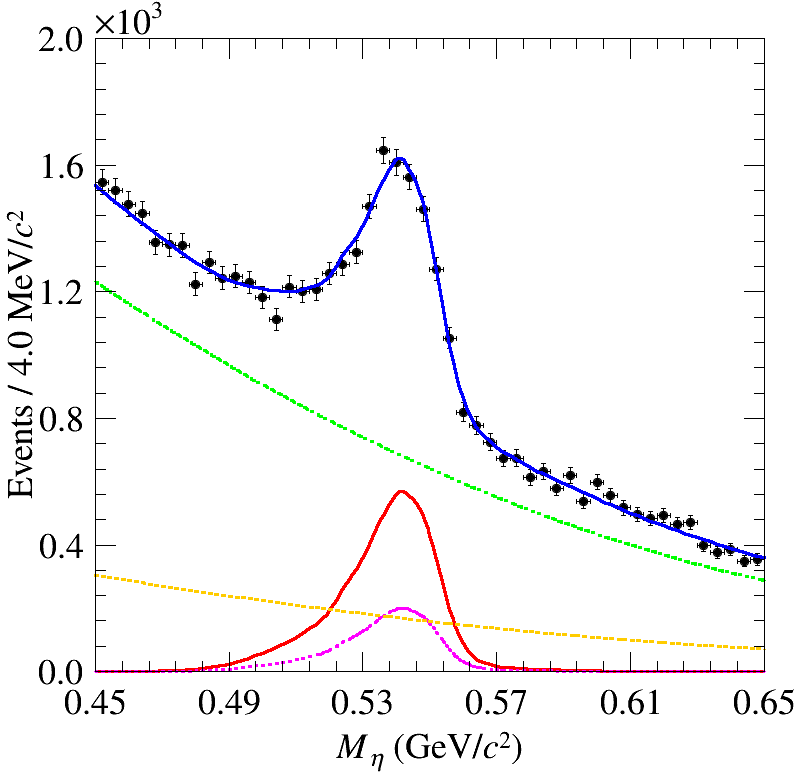}
		\includegraphics[width=7cm]{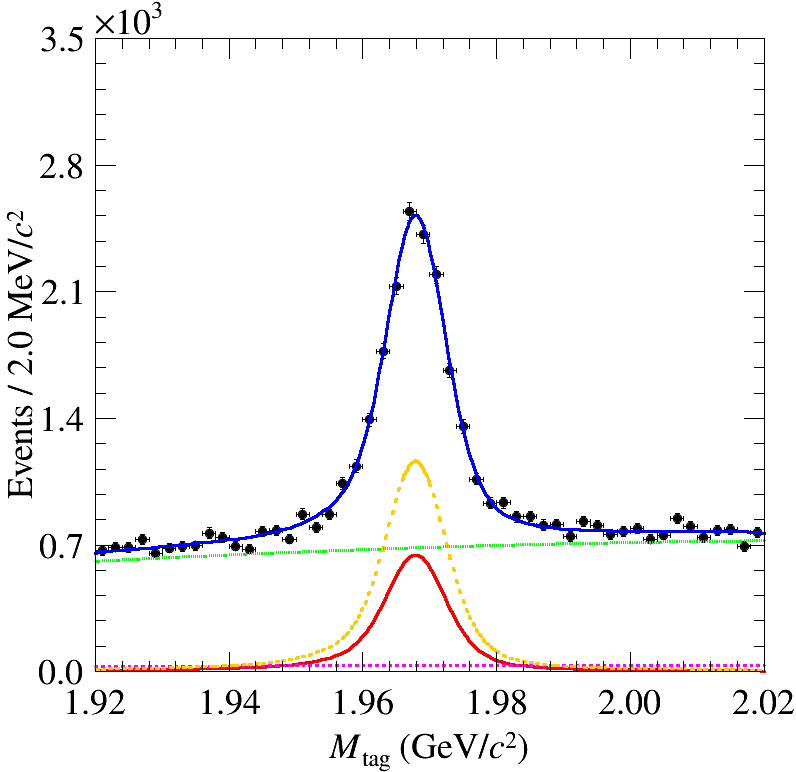}	\\
  		\includegraphics[width=7cm]{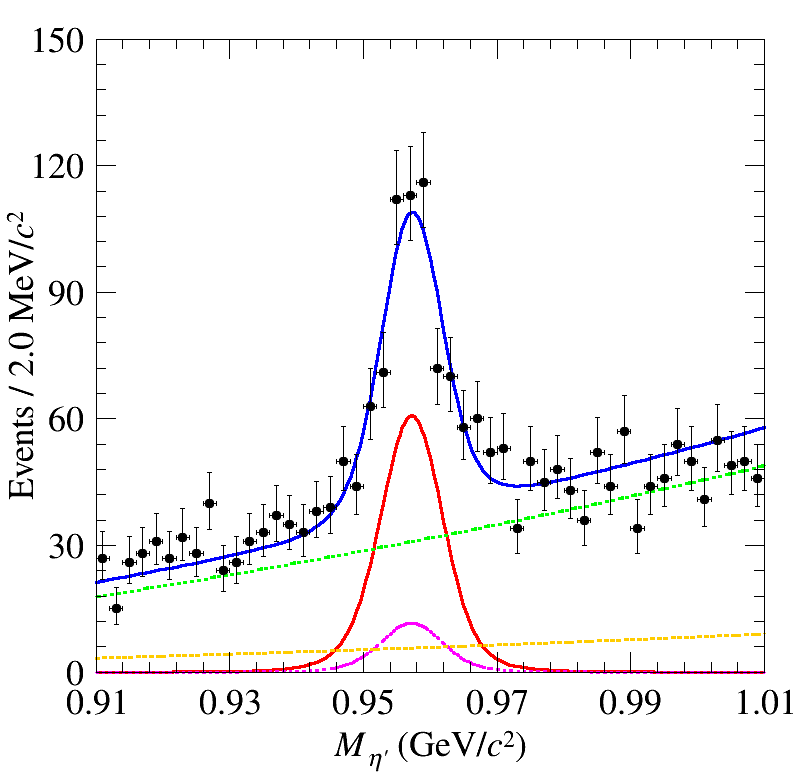}
		\includegraphics[width=7cm]{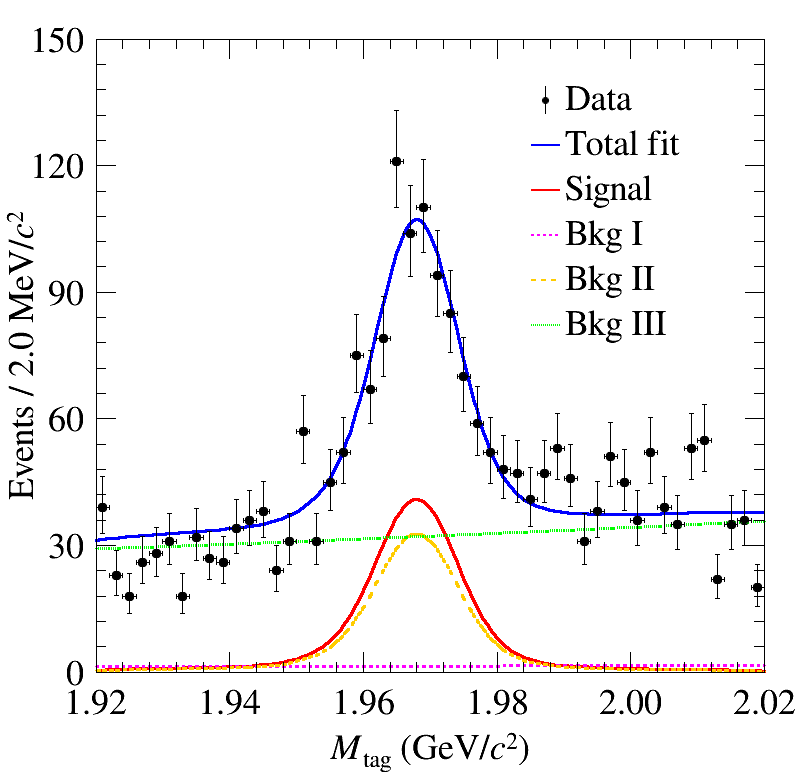}	
\caption{Two-dimensional distributions of $M_{\eta}$ ($M_{\eta^\prime}$) versus
$M_{\rm tag}$. 
Dots with error bars are data. The curves show the
one-dimensional fit projections: solid blue for the total best fit,
solid red for the Signal ($\eta$ ($\eta^\prime$) signal $\times$ tag signal),
magenta dashed for Background I ($\eta$ ($\eta^\prime$) signal $\times$ tag background),
orange short-dashed for Background II (tag signal $\times$ $\eta$ ($\eta^\prime$) background), and
green long-dashed for Background III (background $\times$ background).}
	\label{Fig:FitEta} 
\end{figure*}

The fit results are shown in Fig.~\ref{Fig:FitEta}. From the fits, the DT yields are obtained to be $4726 \pm 136$ and $374 \pm 28$ for $D^{+}_{s} \to \eta X$ and $D^{+}_{s} \to \eta^\prime X$, respectively.  
Their uncertainties are statistical only. The corresponding DT efficiencies \((\epsilon_{\mathrm{tag, sig}}^{\mathrm{DT}})\) are listed in Tables~\ref{Tab:DTEta} and~\ref{Tab:DTEtap}.

%------------------------------------------------

%------------------------------------------------
\begin{table*}[htbp]
	\caption{The DT efficiencies (\%) of the $D^+_s \to \eta X$ decay. The uncertainties are statistical.}
	\begin{tabular}{l r@{$\pm$}l r@{$\pm$}l r@{$\pm$}l r@{$\pm$}l}
		\hline
    Tag mode  &  \multicolumn{2}{c}{$4.128$-$4.157$ GeV} & \multicolumn{2}{c}{$4.178$ GeV} & \multicolumn{2}{c}{$4.189$-$4.219$ GeV} & \multicolumn{2}{c}{$4.226$ GeV}\\
		\hline
		$D^-_s\to K^0_SK^-$          	 & 2.22 & 0.10   & 2.29 & 0.04   & 2.52 & 0.06   & 2.27 & 0.09  \\ 
		$D^-_s\to K^+K^-\pi^-$        	 & 3.63 & 0.06   & 3.66 & 0.02   & 3.78 & 0.03   & 3.45 & 0.05  \\
		$D^-_s\to K^+K^-\pi^-\pi^0$      & 0.68 & 0.04   & 0.78 & 0.02   & 0.77 & 0.02   & 0.77 & 0.04  \\
		$D^-_s\to K^0_SK^+\pi^-\pi^-$    & 1.33 & 0.08   & 1.45 & 0.04   & 1.43 & 0.05   & 1.28 & 0.08  \\
		\hline
	\end{tabular}
	\label{Tab:DTEta}
\end{table*}

%------------------------------------------------
\begin{table*}[htbp]
	\caption{ The DT efficiencies (\%) of the $D^+_s \to \eta^\prime X$ decay. The uncertainties are statistical.}

	\begin{tabular}{l r@{$\pm$}l r@{$\pm$}l r@{$\pm$}l r@{$\pm$}l}
		\hline
    Tag mode  &  \multicolumn{2}{c}{$4.128$-$4.157$ GeV} & \multicolumn{2}{c}{$4.178$ GeV} & \multicolumn{2}{c}{$4.189$-$4.219$ GeV} & \multicolumn{2}{c}{$4.226$ GeV}\\
		\hline
		$D^-_s\to K^0_SK^-$          	  & 1.67 & 0.17   & 1.74 & 0.07   & 1.72 & 0.10   & 1.68 & 0.16  \\
		$D^-_s\to K^+K^-\pi^-$        	  & 2.40 & 0.09   & 2.47 & 0.04   & 2.37 & 0.05   & 2.12 & 0.08  \\
		$D^-_s\to K^+K^-\pi^-\pi^0$       & 1.14 & 0.12   & 1.00 & 0.05   & 0.95 & 0.07   & 1.04 & 0.11  \\
		$D^-_s\to K^0_SK^+\pi^-\pi^-$    & 0.39 & 0.09   & 0.53 & 0.05   & 0.47 & 0.06   & 0.49 & 0.10  \\
		\hline
	\end{tabular}
	\label{Tab:DTEtap}
\end{table*}

%-----------------------------------------------------------------------------
\section{BF measurement}
\label{section:fit}
To determine the BFs, we begin with the equations for an ST mode:
%------------------------------------------------
\begin{equation}
	N^{\rm ST}_{\rm tag}=2N_{D^{*\pm}_sD^{\mp}_s}\mathcal{B}_{\rm tag}\epsilon^{\rm ST}_{\rm tag},
\end{equation}
%------------------------------------------------
\begin{equation}
	N^{\rm DT}_{\rm tag,sig}=2N_{D^{*\pm}_sD^{\mp}_s}\mathcal{B}_{\rm tag}\mathcal{B}_{\rm sig}\epsilon^{\rm DT}_{\rm tag,sig},
\end{equation}
where $N_{D^{*\pm}_sD^{\mp}_s}$ is the total number of $D^{*\pm}_sD^{\mp}_s$ pairs produced from $e^+e^-$ collisions; 
$N^{\rm ST}_{\rm tag}$ and $N^{\rm DT}_{\rm tag,sig}$ are the ST yield  and  the DT yield, respectively; $\mathcal{B}_{\rm tag}$ and $\mathcal{B}_{\rm sig}$ are the BFs of the tag mode and the signal mode, respectively; $\epsilon^{\rm ST}_{\rm tag}$ and $\epsilon^{\rm DT}_{\rm tag,sig}$ are the ST efficiency and the DT efficiency, respectively.
There are more than one tag mode and energy point,
%------------------------------------------------
\begin{equation}
	N^{\rm ST}_{\alpha,i} = 2N_{D^{*\pm}_sD^{\mp}_s}\mathcal{B}_{\alpha}\epsilon^{\rm ST}_{\alpha,i},
\end{equation}
\begin{equation}
	N^{\rm DT}_{\rm total} = \sum_{\alpha,i}{N^{\rm DT}_{\alpha,i,\rm sig}} = \mathcal{B}_{\rm sig}\sum_{\alpha,i}{2N_{D^{*\pm}_sD^{\mp}_s}\mathcal{B}_{\alpha}\epsilon^{\rm DT}_{\alpha,i,\rm sig}},
\end{equation}
where $\alpha$ and $i$ represent the indices for tag modes and energy points, respectively. By isolating $\mathcal{B}_{\rm sig}$, one can obtain
%------------------------------------------------
\begin{equation}
	\mathcal{B}_{\rm sig}=\frac{N^{\rm DT}_{\rm total}}{\sum_{\alpha,i}{N^{\rm ST}_{\alpha,i}}\epsilon^{\rm DT}_{\alpha,i,\rm sig}/\epsilon^{\rm ST}_{\alpha,i}}.
\end{equation}
More precisely, we can compute the BFs of $D^+_s\to\eta X$ and $D^+_s \to \eta^\prime X$ by 
%------------------------------------------------
\begin{equation}
	\mathcal{B}_{D_s^+ \rightarrow \eta X}=\frac{N_{\rm total}^{\rm DT}}{\mathcal{B}_{\eta}\sum_{\alpha,i}{N_{\alpha ,i}^{\rm ST}\epsilon_{\alpha,i,\rm sig}^{\rm DT}/\epsilon_{\alpha ,i}^{\rm ST}}},
	\label{Formular:Eta}
\end{equation}
and
%------------------------------------------------
\begin{equation}
	\mathcal{B}_{D_s^+ \rightarrow \eta^\prime X}=\frac{N_{\rm total}^{\rm DT}}{\mathcal{B}_{\eta^\prime }\mathcal{B}_{\eta}\sum_{\alpha,i} N_{\alpha ,i}^{\rm ST}\epsilon_{\alpha,i,\rm sig}^{\rm DT}/\epsilon_{\alpha ,i}^{\rm ST}}, 
	\label{Formular:Etap}
\end{equation}
where $\mathcal{B}_{\eta^{\prime}}$ and $\mathcal{B}_{\eta}$ are the  known BFs of $\eta^{\prime} \to \pi^{+}\pi^{-}\eta$ and $\eta \to \gamma\gamma$~\cite{PDG}, respectively. 
Using Eq.~(\ref{Formular:Eta}) and Eq.~(\ref{Formular:Etap}) together with the ST yields from Table~\ref{Tab:ST}, ST efficiencies from Table~\ref{Tab:STeff}, DT efficiencies from Tables~\ref{Tab:DTEta} and \ref{Tab:DTEtap}, and the obtained DT yields, we determine the BFs of $D^{+}_{s} \to \eta X$ and $D^{+}_{s} \to \eta^\prime X$ to be $(30.72 \pm 0.90)\%$ and $(12.30 \pm 0.94)\%$, respectively, where the uncertainties are statistical only.

To validate the reliability of the measurement procedure, we perform closure checks on the measured BFs by using forty inclusive MC samples, each of which  corresponds to the same integrated luminosity of the real data. We  find that the measured BFs for each sample are consistent with the input values  and their pull distributions obey a normal distribution, thereby ensuring the reliability of our analysis procedure. 
%------------------------------------------------------------------------------------------------------------
\section{Systematic uncertainties}
\label{section:sys}

Benefiting from the DT method, the uncertainties related to the tag side largely cancel. 
The  following sources of systematic uncertainties are taken into account.

%------------------------------------------------
\begin{itemize}
\item \textbf{$\pi^{+}$ tracking: }
  The $e^+e^- \to K^+K^-\pi^+\pi^-~(\pi^0)$, and $ \pi^+\pi^-\pi^+\pi^-~(\pi^0)$ processes are used to study the  $\pi^+$ tracking efficiency. 
The systematic uncertainty for tracking each charged pion is assigned as 1.0\%.
 \item \textbf{$\pi^{+}$ PID: }
The $\pi^{+}$ PID efficiency is  studied with the same control samples used for the tracking study.  
The systematic uncertainty for PID of each charged pion is assigned as 0.5\%.

 \item \textbf{$\eta$ reconstruction: }
The systematic uncertainty associated with the $\eta$ reconstruction (including the reconstruction of the two photons) is determined to be 0.5\%, based on the study of the control sample of $e^+e^- \to K^+K^-\pi^+\pi^-\pi^0$~\cite{PhysRevD.99.091101}, owing to the limited $\eta$ sample.

  \item\textbf{Radiative $\gamma$ reconstruction:} 
The systematic uncertainty of the radiative $\gamma$ construction from $D^{*+}_s$ is assigned to be $1.0\%$ per photon, by studying the control sample of the ${\mathit J / \psi} \to \pi^+\pi^-\pi^0$~\cite{sys_gamma}.
 
	\item \textbf{ST $D_s^-$ yield}: 
The uncertainty from the ST $D_s^-$ yield is determined as a 0.3\% systematic uncertainty from varying the background shape in the ST fit.
		
  	\item \textbf{2D fit}: 
The uncertainties arising from the signal and background models  are evaluated by varying the fit shapes. 
The systematic uncertainty from the signal model is assessed by substituting the nominal $M_{\eta}$ ($M_{\eta^{\prime}}$) and $M_{\rm tag}$ shapes with their MC-simulated counterparts (without Gaussian convolution), yielding uncertainties of 0.7\% (0.4\%) and 0.4\% (0.3\%), respectively. For the background model,  the similar alternative background models cause uncertainties of 0.2\% (0.8\%) and 0.4\% (0.3\%), respectively.

	\item \textbf{MC statistics}:
The uncertainty due to limited sample sizes is estimated to be 0.3\% (0.2\%) for $D_s^+ \to \eta X$ ($D_s^+ \to \eta^\prime X$), obtained by propagating the statistical uncertainties of the ST efficiencies at different energy points according to $\sqrt{\sum_{\alpha,i} \left( f_{\alpha,i} \cdot \frac{\delta_{\epsilon^{\rm ST}_{\alpha,i}}}{\epsilon^{\rm ST}_{\alpha,i}} \right)^2}$, where $f_{\alpha,i}$ is the ST yield fraction and $\delta_{\epsilon^{\rm ST}_{\alpha,i}}$ is the statistical uncertainty of the ST efficiency for the $i$-th tag mode at the $\alpha$-th energy point.

	\item \textbf{Quoted BFs}:
		We account for the systematic uncertainties arising from the quoted BFs of $\eta^\prime \to \pi^+ \pi^- \eta$, $\eta \to \gamma\gamma$ and $D_s^{*+} \to \gamma D_s^+$ decays~\cite{PDG}. The relative uncertainties assigned are 0.6\% for the $\eta \to \gamma\gamma$ decay and 1.3\% for the $\eta^\prime \to \pi^+ \pi^- \eta$ decay.
\end{itemize}

All systematic uncertainties discussed above are summarized in Table~\ref{Tab:Systematic}. 
The total systematic uncertainty is obtained by adding these uncertainties in quadrature.

%------------------------------------------------
\begin{table}[htbp]
	\caption{The relative systematic uncertainties ($\%$) in the inclusive BF measurements.}

	\begin{tabular}{lcc}
		\hline
		Source   									&$D^+_s\to\eta X$ 	&$D^+_s\to\eta^\prime X$ 	\\
		\hline
		Tracking                           			&-     		            		&2.0									\\ 
		PID                             			&-              				&1.0									\\          
		$\eta$ reconstruction        				&0.5   							&0.5									\\     
		Radiative $\gamma$ reconstruction        				&1.0   							&1.0									\\
		ST $M_{D_s^-}$						&0.3   							&0.3									\\ 
		$M_{\eta}$ ($M_{\eta^\prime}$) signal shape		&0.7   							&0.4   									\\ 
		$M_{\eta}$ ($M_{\eta^\prime}$) background shape	&0.2      						&0.8									\\  
		MC statistics								&0.3  	 						&0.2									\\ 
		Quoted BFs 									&0.6   							&1.3									\\
		\hline
		Total										&1.5							&3.0									\\
  		\hline
	\end{tabular}
	\label{Tab:Systematic}
\end{table}

%-----------------------------------------------------------------------------

%-------------------------------------------

%-----------------------------------------------------------------------------
%------------------------------------------------

\begin{table*}[htbp]
	\caption{Comparison  of the BFs (\%) between the previous measurements  and this work. The first and second uncertainties are statistical and systematic, respectively.}
	\begin{tabular}{l  r@{@}l c r@{$\pm$}c@{$\pm$}l r@{$\pm$}c@{$\pm$}l}
		\hline
		Experiment    					&  \multicolumn{2}{c}{${\cal L}_{\rm int}$ and $\sqrt s$}   & Process			&\multicolumn{3}{c}{$\mathcal{B}(D^+_s\to\eta X)$}	&\multicolumn{3}{c}{$\mathcal{B}(D^+_s\to\eta^\prime X)$}	\\             
		\hline
		CLEO~\cite{Ref:CLEO2009}  	& ~586 pb$^{-1}$ & $\sqrt{s}=4.170$ GeV &	~$D_s^{*\pm}D^\mp_s$	&~$29.90 $ & $2.20 $ & $1.70$    &~$11.70 $ & $1.70 $ & $0.70$      	 \\
		BESIII~\cite{BESIII:2015rrp} & ~482 pb$^{-1}$ & $\sqrt{s}=4.009$ GeV  & 	~$D_s^+D^-_s$	&\multicolumn{3}{c}{-}                             &~$8.80 $ & $1.80 $ & $0.50$      	\\
		This work  				 &  ~7.33 fb$^{-1}$ & $\sqrt{s}=4.128$-$4.226$ GeV& ~$D_s^{*\pm}D^\mp_s$ & ~$30.72 $ & $  0.89 $ & $ 0.49$ & ~$12.30 $ & $  0.94 $ & $ 0.37$  		\\
		\hline
	\end{tabular}
	\label{Tab:ResultCompare}
\end{table*}

\section{Summary}

The BFs of the inclusive decays $D_s^+ \to \eta X$ and $D_s^+ \to \eta^\prime X$ are measured using $7.33~\text{fb}^{-1}$ of $e^+ e^-$ collision data collected at center-of-mass energies between $4.128$ and $4.226~\text{GeV}$ with the BESIII detector. The obtained results are $\mathcal{B}(D_s^+ \to \eta X) = (30.72 \pm 0.90 \pm 0.46)\%$ and $\mathcal{B}(D_s^+ \to \eta^\prime X) = (12.30 \pm 0.94 \pm 0.37)\%$. 
A comparison with previous measurements is provided in Table~\ref{Tab:ResultCompare}. 
The measured BFs for $D_s^+ \to \eta X$ and $D_s^+ \to \eta^\prime X$ are consistent with the measurements from the CLEO experiment~\cite{Ref:CLEO2009}. 
The precision of our BF for the $D_s^+ \to \eta X$ ($D_s^+ \to \eta^\prime X$) decay is $3.3\%$ ($8.2\%$), corresponding to approximately 2.8 (1.9) times improvement compared to the CLEO measurements~\cite{Ref:CLEO2009}.
These precise measurements not only improve the understanding of $D_s^+$ decay dynamics but also serve as essential inputs for normalizing $B$ and $B_s$ decays, constraining systematic uncertainties, and improving background modeling in future heavy-flavor physics studies.

Comparing our BFs of $D^+_s\to \eta X$ and  $D^+_s\to \eta^\prime X$  with the sums of the currently known exclusive BFs listed in Tables~\ref{Tab:BF1} and  \ref{Tab:BF2}, respectively, gives the difference:
\begin{equation}
\mathcal{B}(D_s^+ \to \eta X)-\sum_i \mathcal{B}(D_s^+ \to \eta X_i)=(5.33 \pm 1.04)\%
\nonumber
\end{equation}
\begin{equation}
\mathcal{B}(D_s^+ \to \eta^\prime X)-\sum_i \mathcal{B}(D_s^+ \to \eta^\prime X_i)=(0.33 \pm 1.03)\%
\nonumber
\end{equation}
where $\sum_i \mathcal{B}(D_s^+ \to \eta X_i) = (25.40 \pm 0.20)\%$ and $\sum_i \mathcal{B}(D_s^+ \to \eta^\prime X_i) = (11.97 \pm 0.21)\%$ are obtained from the PDG~\cite{PDG}. 
Our measurements suggest possible unobserved $D_s^+$ decay modes involving an $\eta$ meson, while no significant excess is observed for modes containing an $\eta^\prime$ meson, thereby constraining the total BFs of all unobserved decays in exclusive  $D_s^+ \to \eta^{(\prime)} X$ decays.

%------------------------------------------------------------------------------

\begin{acknowledgments}

The BESIII Collaboration thanks the staff of BEPCII (https://cstr.cn/31109.02.BEPC) and the IHEP computing center for their strong support. This work is supported in part by National Key R\&D Program of China under Contracts Nos. 2023YFA1606000, 2023YFA1606704, 2025YFA1613900; National Natural Science Foundation of China (NSFC) under Contracts Nos. 11635010, 11935015, 11935016, 11935018, 12025502, 12035009, 12035013, 12061131003, 12192260, 12192261, 12192262, 12192263, 12192264, 12192265, 12221005, 12225509, 12235017, 12342502, 12361141819, 12535005; the Chinese Academy of Sciences (CAS) Large-Scale Scientific Facility Program; the Strategic Priority Research Program of Chinese Academy of Sciences under Contract No. XDA0480600; CAS under Contract No. YSBR-101; 100 Talents Program of CAS; The Institute of Nuclear and Particle Physics (INPAC) and Shanghai Key Laboratory for Particle Physics and Cosmology; Agencia Nacional de Investigaci\'on y Desarrollo de Chile (ANID), Chile under Contract No. ANID CCTVal CIA250027; ERC under Contract No. 758462; German Research Foundation DFG under Contract No. FOR5327; Istituto Nazionale di Fisica Nucleare, Italy; Knut and Alice Wallenberg Foundation under Contracts Nos. 2021.0174, 2021.0299, 2023.0315; Ministry of Development of Turkey under Contract No. DPT2006K-120470; National Research Foundation of Korea under Contract No. RS-2026-25486791; National Science and Technology fund of Mongolia; Polish National Science Centre under Contract No. 2024/53/B/ST2/00975; STFC (United Kingdom); Swedish Research Council under Contract No. 2019.04595; U. S. Department of Energy under Contract No. DE-FG02-05ER41374
\end{acknowledgments}

\bibliographystyle{apsrev4-2} 
\bibliography{ref.bib}

\begin{widetext}
\centering
%% Saved at => 2026-05-08
M.~Ablikim$^{1}$\BESIIIorcid{0000-0002-3935-619X},
M.~N.~Achasov$^{4,c}$\BESIIIorcid{0000-0002-9400-8622},
P.~Adlarson$^{84}$\BESIIIorcid{0000-0001-6280-3851},
X.~C.~Ai$^{90}$\BESIIIorcid{0000-0003-3856-2415},
C.~S.~Akondi$^{32A,32B}$\BESIIIorcid{0000-0001-6303-5217},
R.~Aliberti$^{40}$\BESIIIorcid{0000-0003-3500-4012},
A.~Amoroso$^{83A,83C}$\BESIIIorcid{0000-0002-3095-8610},
Q.~An$^{79,66,\dagger}$,
Y.~H.~An$^{90}$\BESIIIorcid{0009-0008-3419-0849},
M.~S.~Anderson$^{40}$\BESIIIorcid{0009-0008-1550-2632}
% ...
\\
\vspace{0.2cm}
(BESIII Collaboration)\\
\vspace{0.2cm} {\it
$^{1}$ Institute of High Energy Physics, Beijing 100049, People's Republic of China\\
% ...
$^{4}$ Budker Institute of Nuclear Physics SB RAS (BINP), Novosibirsk 630090, Russia\\
% ...
$^{32}$ INFN Sezione di Ferrara, (A)INFN Sezione di Ferrara, I-44122, Ferrara, Italy; (B)University of Ferrara, I-44122, Ferrara, Italy\\
% ...
$^{40}$ Johannes Gutenberg University of Mainz, Johann-Joachim-Becher-Weg 45, D-55099 Mainz, Germany\\
% ...
$^{66}$ State Key Laboratory of Particle Detection and Electronics, Beijing 100049, Hefei 230026, People's Republic of China\\
% ...
$^{79}$ University of Science and Technology of China, Hefei 230026, People's Republic of China\\
% ...
$^{83}$ University of Turin and INFN, (A)University of Turin, I-10125, Turin, Italy; (B)University of Eastern Piedmont, I-15121, Alessandria, Italy; (C)INFN, I-10125, Turin, Italy\\
$^{84}$ Uppsala University, Box 516, SE-75120 Uppsala, Sweden\\
% ...
$^{90}$ Zhengzhou University, Zhengzhou 450001, People's Republic of China\\

\vspace{0.2cm}
$^{\dagger}$ Deceased\\
% ...
$^{c}$ Also at the Novosibirsk State University, Novosibirsk, 630090, Russia\\
% ...

}
%% ends here %%

\end{widetext}
\end{document}